\documentclass[12pt, a4paper]{iopart}

\usepackage[a4paper]{geometry}
\usepackage[T1]{fontenc}\usepackage[latin1]{inputenc}
\usepackage{graphicx,color,booktabs,wrapfig,microtype,sidecap,floatflt,afterpage,cite,hvfloat}
\renewcommand{\figurename}{Figure}

\begin{document}
\makeatletter
\def\@evenhead{\hfil\itshape\rightmark}
\def\@oddhead{\itshape\leftmark\hfil}
\renewcommand{\@evenfoot}{\hfill\small\raisebox{0em}{--~\textbf{\thepage}~--}\hfill}
\renewcommand{\@oddfoot}{\hfill\small\raisebox{0em}{--~\textbf{\thepage}~--}\hfill}
\makeatother

\pagestyle{plain}
	
\title{Magnetic phase diagram of the helimagnetic spinel compound ZnCr$_{2}$Se$_{4}$ revisited by small-angle neutron scattering}

\author{A.~S.~Cameron$^{1}$, Y.~V.~Tymoshenko$^{1}$, P.~Y.~Portnichenko$^{1}$, J.~Gavilano$^{2}$, V.~Tsurkan$^{3,4}$, V. Felea$^{4}$, A. Loidl$^{3}$, S. Zherlitsyn$^{5}$, J. Wosnitza$^{1,5}$, D.~S.~Inosov$^{1}$}
\address{$^{1}$Institut f\"ur Festk\"orperphysik, TU Dresden, D-01069 Dresden, Germany}
\address{$^{2}$Laboratory for Neutron Scattering, Paul Scherrer Institut, 5232 Villigen PSI, Switzerland}
\address{$^{3}$Experimental Physics V, Center for Electronic Correlations and Magnetism, University of Augsburg, D-86135 Augsburg, Germany}
\address{$^{4}$Institute of Applied Physics, Academy of Sciences of Moldova, MD 2028 Chisinau, Republic of Moldova}
\address{$^{5}$Hochfeld-Magnetlabor Dresden (HLD-EMFL), Helmholtz-Zentrum Dresden-Rossendorf, D-01314 Dresden, Germany}

\begin{abstract}
We performed small-angle neutron scattering (SANS) measurements on the helimagnetic spinel compound ZnCr$_{2}$Se$_{4}$. The ground state of this material is a multi-domain spin-spiral phase, which undergoes domain selection in a magnetic field and reportedly exhibits a transition to a proposed spin-nematic phase at higher fields. We observed a continuous change in the magnetic structure as a function of field and temperature, as well as a weak discontinuous jump in the spiral pitch across the domain-selection transition upon increasing field. From our SANS results we have established the absence of any long-range magnetic order in the high-field (spin-nematic) phase. We also found that all the observed phase transitions are surprisingly isotropic with respect to the field direction.
\end{abstract}

%\submitto{\JPCM}
\maketitle

\section*{Introduction}

ZnCr$_{2}$Se$_{4}$ is a magnetoelectric compound, possessing the cubic spinel ($Fd\bar{3}m$) structure in the paramagnetic phase with a lattice parameter of $a = 10.497$ \AA \cite{Kle66, Plu66}. Magnetoelectric compounds are of particular interest as they allow for a possible realisation of devices with mutual magnetic and electric control, and the nature of the coupling between the magnetisation and electric polarisation in these materials is not yet fully understood. Along with a number of other isostructural compounds, ZnCr$_{2}$Se$_{4}$ has demonstrated a linear magnetoelectric effect, in particular a linear magnetoelectric effect for electric fields applied perpendicular to the propagation vector of its spin-spiral ground state \cite{Sir80a, Mur08}.

Early neutron diffraction measurements have shown that in the absence of an applied field the Cr$^{3+}$ $S=3/2$ moments in ZnCr$_{2}$Se$_{4}$ form an incommensurate helical structure propagating along [100]. A variety of spin-screw structures are observed in helimagnetic compounds, and a comprehensive discussion of these spin arrangements in various materials was the subject of several reviews \cite{Tok10, Kim12}. Our sample magnetically orders at $T_{\mathrm{N}} \approx 21$~K \cite{Fel12}, and at low temperature in zero field the incommensurate screwlike magnetic structure has a reported helical pitch of 22.4 \AA, which is equivalent to 6.1 Cr--Cr distances and corresponds to an angle of 42$^{\circ}$ between spins in consecutive ferromagnetic planes orthogonal to the spiral \cite{Plu66}. While at high temperatures, above $T_{\rm N}$, the crystal structure of ZnCr$_{2}$Se$_{4}$ is cubic, upon crossing the magnetic ordering transition the crystal is seen to undergo a distortion which is concurrent with the magnetic transition \cite{Kle66}. Initial measurements indicated that this was a cubic to tetragonal distortion with a $c/a$ ratio of around 0.999 \cite{Kle66}. Later measurements, however, discovered that the crystal symmetry of the ground state is actually orthorhombic, where $a \cong b > c$ \cite{Hid03, Yok09}.  The concomitant structural and magnetic phase transitions would suggest that the structural distortion is magnetically driven, which is supported by the observation of a spin-lattice coupling at $T_{\mathrm{N}}$ by ultrasound measurements \cite{Fel12} as well as by neutron diffraction \cite{Plu66, Aki78}. It is seen that the spin-spiral structure propagates along the crystal axis which experiences the maximum distortion at the structural transition \cite{Kle66, Hid03}, which is the \textbf{c} axis in the previous definition. There are three possible domains, owing to the cubic crystal symmetry at high temperatures, and in a small magnetic field, domain selection occurs \cite{Yok09, Fel12}. Measurements on a single crystal show that there is usually one dominant domain, although there may be a minor fraction of the other two possible domains present in the crystal, and the relative population of these domains exhibits a hysteresis across the magnetic transition line \cite{Kle66, Hid03}.

At low temperature, the spin-spiral state persists until 6~T, although in applied field the magnetic structure is transformed, at $H_{\rm C1}$, into a longitudinal conical spiral, as the opening angle is reduced with increasing field \cite{Fel12, Plu66, Mur08}. At 6~T, the lattice is nearly ferromagnetically polarised, and above this field the system gives way to a new high-field phase which is proposed to be a spin-nematic state \cite{Fel12}. Here, the rate of polarisation with field is greatly reduced in contrast to the spin-spiral state, yet the spin lattice continues to polarise with increasing field until 10~T, where the system then enters the fully saturated ferromagnetic state \cite{Fel12}.

The proposed spin-nematic phase is of great interest, as it remains uncharacterised in this compound. In particular, the arrangement of the unsaturated components of the magnetic moments in direct space remains a puzzle. The possibility of a spin-nematic state arising in a helimagnetic system is a long-standing prediction \cite{Cha90}, and such a scenario has been identified in LiCuVO$_{4}$ \cite{Svi11}. The field dependence of the magnetisation in the proposed spin-nematic state of ZnCr$_{2}$Se$_{4}$ \cite{Fel12} is analogous to that of LiCuVO$_{4}$ \cite{Svi11}, where a shallow plateaux preceding the ferromagnetic state is also seen. Using neutron scattering we search for signs of magnetic ordering within this phase to shed light on its character and demonstrate that the unpolarised component of the spins above $H_{\rm C2}$ remains fully disordered without forming any long- or short-range spiral order.

\section*{Experimental method}

We performed small-angle neutron scattering (SANS) measurements on a coaligned assembly of ZnCr$_{2}$Se$_{4}$ single crystals. Crystal preparation was described previously in \cite{Fel12}. The experiment was performed with the magnetic field applied horizontally, i.e. perpendicular to the neutron beam. We used two orientations of the sample with either the [100] or [110] crystallographic direction pointing along the field, whereas the [001] direction for zero rocking angle was pointing along the neutron beam in both cases. These two configurations correspond to the field orientation along the natural propagation vector of the magnetic structure and halfway between two such propagation vectors in neighbouring magnetic domains, respectively. For the field parallel to [100], the sample mass was approximately one gram, consisting of six crystals co-aligned with X-ray Laue diffraction and mounted on an aluminium plate. For the field applied parallel to [110], only four crystals from the mosaic were used. These were placed within a cryomagnet with a base temperature of around 2~K and a maximum field of 11~T. Neutron diffraction measurements were performed on the SANS-I instrument at the Paul Scherrer Institute, with the incoming neutron wavelength set to 4.7~\AA. The sample together with the magnetic field was then rocked over the full range of angles corresponding to the accessible Bragg reflections, with background measurements taken above $T_{\rm N}$ to eliminate all nonmagnetic contributions to the signal. All measurements were taken after cooling the sample to base temperature in zero field, and then applying the required magnetic field. The exception to this is a scan at base temperature which was taken in decreasing field and a following scan in temperature. These exceptions are indicated where applicable.

\section*{Results}

\begin{figure*}
	\includegraphics[width=1\linewidth]{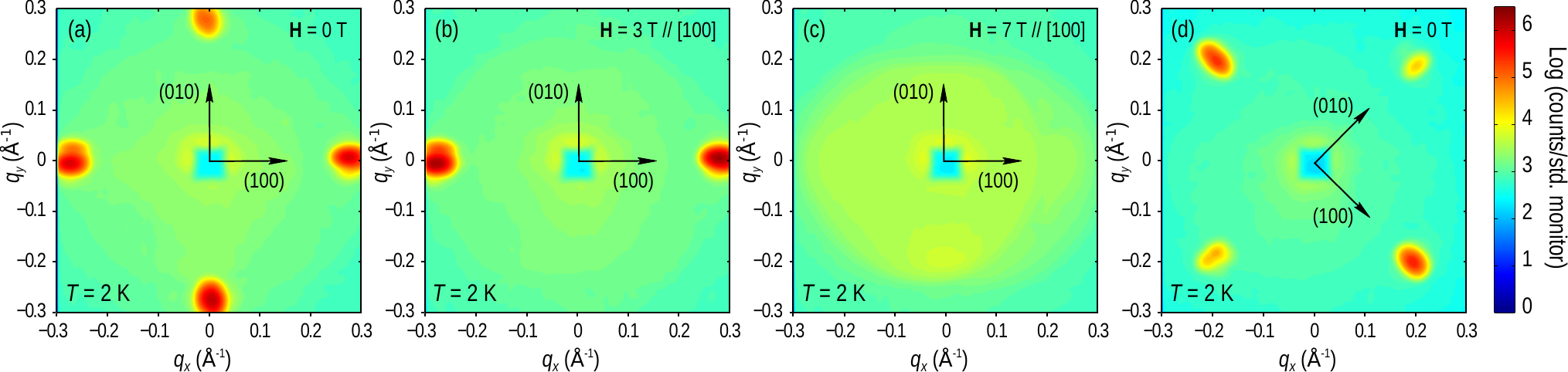}
	\caption{(Colour online) Diffraction patterns from the magnetic structure of ZnCr$_{2}$Se$_{4}$, measured at a base temperature of 2~K for several magnetic fields. (a -- c) Diffraction patterns for the magnetic field aligned along the [100] direction, in applied fields of (a) 0~T, (b) 3~T and (c) 7~T. (d) Diffraction pattern at 0~T for the realigned sample so that the magnetic field could be applied along the [110] direction. Diffraction patterns are a sum over all rocking angles corresponding to the observed Bragg reflections, presented on a logarithmic intensity scale, and the axes $q_{x}$ and $q_{y}$ are in the laboratory frame.}
		\label{diff_pattern}
		\vspace{-2em}
\end{figure*}

Figure~\ref{diff_pattern} shows a selection of representative neutron diffraction patterns from each of the distinct states and orientations observed during the experiment at the base temperature of 2~K. Each image is a sum over rocking angles in the vicinity of the Bragg condition for each individual peak, set to a logarithmic intensity scale. Panels (a -- c) show diffraction patterns from the first orientation, where the magnetic field was applied parallel to the [100] direction, which is horizontal in these figures. The diffraction pattern in panel (a) was taken in zero field, and we can clearly see the two sets of Bragg reflections resulting from the domains aligned along the [010] and [100] directions, which in this image are vertical and horizontal, respectively. The third set of Bragg peaks, along the one remaining [001] direction, are not accessible within this experimental geometry. Panel (b) is taken in an applied field of 3~T, and it can be seen that domain selection has taken place as only one set of Bragg peaks remains, which belong to the domain whose propagation vector is parallel to the magnetic field. Panel (c) was taken in an applied field of 7~T. This is above $H_{\rm C2}$, within the proposed spin-nematic phase, and no magnetic signal can be seen in this measurement. Panel (d) shows the diffraction pattern observed in the second experimental configuration, where the sample was remounted in such a way that the field was also applied horizontally but the crystal axes have been rotated by 45$^{\circ}$ such that the field is applied along [110], equidistant from the propagation vectors of the two magnetic domains. This diffraction pattern was observed at all fields, throughout the long-range ordered phase below $H_{\rm C2}$, indicating that the propagation vector is insensitive to the direction of magnetic field. In this orientation the domain-selection transition at $H_{\rm C1}$, was still
\begin{wrapfigure}{r}{0.5\textwidth}
	\includegraphics[width=1\linewidth]{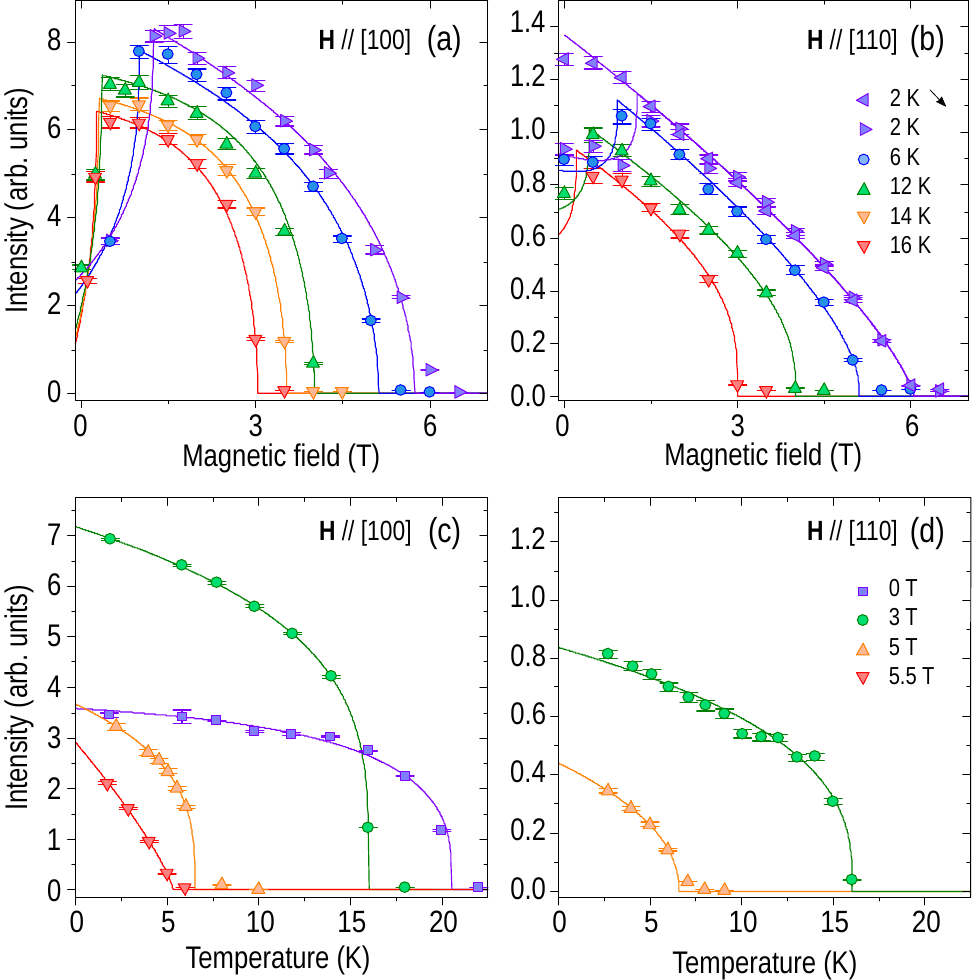}
	\caption{(Colour online) Dependence of the magnetic Bragg intensity on the applied field and sample temperature. (a,~b) Intensity of the Bragg reflections as a function of magnetic field applied along (a) the [100] direction and (b) the [110] direction. The arrow in the legend of panel~(b) indicates the single scan done in decreasing field. (c,~d) Intensity of the Bragg reflections as a function of temperature for magnetic fields applied along (c) [100] and (d) [110], after zero-field cooling. The lines are guides for the eyes.}
		\label{intensity}
		\vspace{-2.2em}
\end{wrapfigure}
observed as an increase in the Bragg peak intensity of the two equivalent domains at the expense of the third domain whose propagation vector is orthogonal to the applied magnetic field in figure~\ref{diff_pattern}(d).

Figure~\ref{intensity} presents the intensity of the Bragg reflections within the domains which are preferentially selected at $H_{\rm C1}$, illustrated in figure~\ref{diff_pattern} as a function of both field and temperature, for magnetic field orientations along both [100] and [110]. The data in panel~(a) represent the intensity of the Bragg reflections as a function of magnetic field, applied after zero-field cooling along [100]. Here, we see that there is a rapid increase in Bragg intensity at low magnetic field, in the same region where the domain selection transition takes place. Panel (b) shows the corresponding intensity of the Bragg reflections as a function of magnetic field applied along the [110] direction. It shows the same behaviour as for the field applied along [100] in panel~(a), however the magnitude of the increase in Bragg intensity as a result of domain selection is approximately twice smaller. This is fully consistent with the number of domains selected for each field orientation, which is 1 out of 3 for $\textbf{H} \parallel [100]$ and 2 out of 3 for $\textbf{H} \parallel [110]$. These measurements were also taken after zero-field cooling, except for one scan at 2~K shown in figure~\ref{intensity}(b), done with decreasing field. This measurement shows no decrease of intensity across the domain selection transition, and instead the signal strength increases continuously down to zero field, indicating that the domain distribution is not affected by the removal of magnetic field. After accounting for the domain selection, for all measurements we observe a decrease in Bragg intensity as a function of increasing magnetic field until it reaches zero at $H_{\rm C2}$, with no Bragg scattering observed between $H_{\rm C2}$ and $H_{\rm C3}$ within the covered region of momentum space. Panel~(c) shows the intensity as a function of temperature for magnetic fields of 0, 3, 5 and 5.5~T applied along [100]. All curves show an order-parameter-like monotonic decrease, with a sharp fall-off in intensity at $T_{\rm N}$. Panel~(d) displays the diffracted intensity of the Bragg reflections for magnetic fields applied along the [110] direction. This shows the same general behaviour as for fields applied along [100], where the intensity falls off with increasing temperature and reaches zero at $T_{\rm N}$.

\begin{wrapfigure}{r}{0.5\textwidth}\vspace{-1.0em}
	\includegraphics[width=1\linewidth]{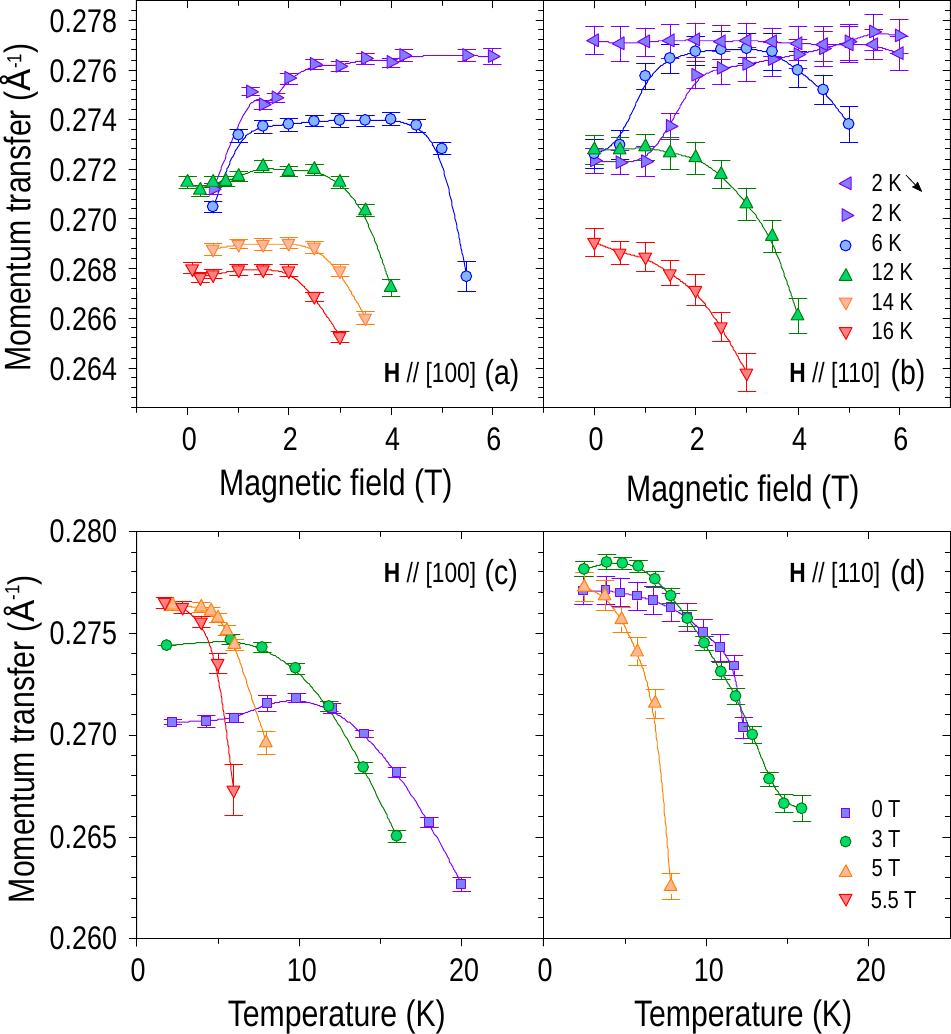}
	\caption{(Colour online) The scattering vector of the Bragg reflections from the magnetic structure as a function of both field and temperature. (a) Scattering vector as a function of magnetic field applied along the [100] direction. (b) The same for magnetic field applied along the [110] direction. (c) Scattering vector as a function of temperature for magnetic field applied along the [100] direction. (d) The same for magnetic field applied along the [110] direction. The lines are a guide for the eyes.}
		\label{q}
		\vspace{-1.5em}
\end{wrapfigure}

Figure~\ref{q} shows the propagation vector of the magnetic structure, $\vert \textbf{q} \vert$, as a function of both applied magnetic field and temperature. Panels (a) and (b) depict $\vert \textbf{q} \vert$ as a function of magnetic field applied along the [100] and [110] directions, respectively, at temperatures of 2, 6, 12, 14 and 16~K. Generally, the changes in $\vert \textbf{q} \vert$ are weak, on the order of a few percent. The most striking feature of the data in both of these panels is the rapid increase in the propagation vector at low fields seen in the low-temperature measurements. This occurs between 1 and 2~T at 2~K, and it is clear from the diffraction patterns that this change in $\vert \textbf{q} \vert$ coincides with the domain-selection transition as the Bragg spots from domains perpendicular to the magnetic field disappear. The location of this transition in magnetic field is suppressed with increasing temperature, and so it is not seen at higher temperatures. Panel (b) also displays the scan at 2~K in decreasing fields, which expectedly shows no rapid change in $\vert \textbf{q} \vert$, since domain selection only occurs when applying a magnetic field after zero-field cooling, whereas the removal of magnetic field at base temperature does not alter the domain distribution. We see that in both orientations, at higher temperatures, there is a slight decrease in $\vert \textbf{q} \vert$ when approaching the phase transition at $H_{\rm C2}$.

Panels (c) and (d) show $\vert \textbf{q} \vert$ as a function of increasing temperature for the same two  magnetic-field directions. All these measurements were taken after cooling the sample in zero field, except for the 0~T measurement in panel (d), as this was taken after the decreasing field scan at 2~K shown in panel (b), in order to remain in the domain-selected state for consistency with the other curves. As a result of this different field history, the shallow maximum seen in the 0~T scan in panel (c), where the magnetic field was applied along [100] after zero-field cooling, is not observed in the same measurement for the field applied along [110] in panel (d), where the domain selection transition was deliberately circumvented.

\begin{wrapfigure}{r}{0.5\textwidth}\vspace{-1.3em}
	\includegraphics[width=1\linewidth]{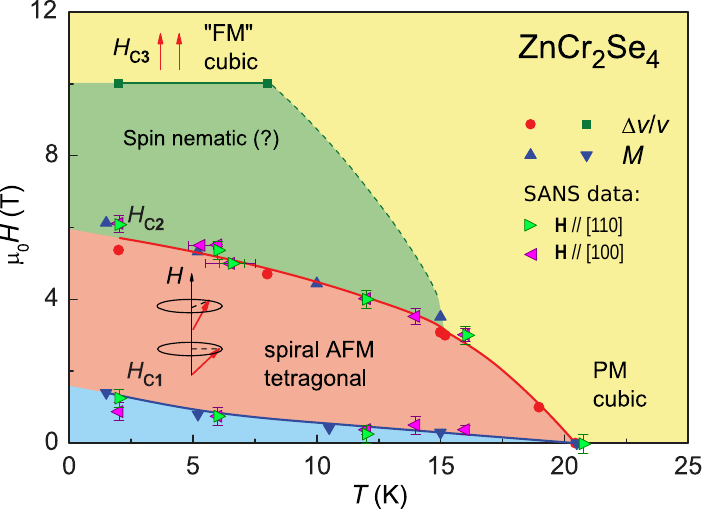}
	\caption{(Colour online) Applied magnetic field vs. temperature phase diagram for ZnCr$_{2}$Se$_{4}$, reproduced from \cite{Fel12}. Here, we superimpose the phase transition and domain selection points deduced from our SANS data onto the original graph constructed from magnetisation (\textit{M}) and sound velocity ($\Delta v/v$) measurements.}
	\vspace{-5em}
		\label{phase_diagram}
\end{wrapfigure}

Finally, figure~\ref{phase_diagram} shows the magnetic field and temperature phase diagram for ZnCr$_{2}$Se$_{4}$, reproduced from \cite{Fel12}. We have superimposed the data points indicating either domain selection or magnetic ordering transition that resulted from the fits to out field- and temperature-dependent SANS measurements onto this diagram. To within experimental uncertainty our data for both field directions coincide and show excellent agreement with the previous measurements.

\section*{Discussion}

Our neutron-scattering measurements revealed that the magnetic signal from the spin-spiral structure always vanishes at the transition line $H_{\rm C2}$, regardless of whether the system is undergoing a transition to the paramagnetic state or into the proposed spin-nematic state. We observed no other Bragg scattering within the high-field phase. We can, therefore, establish the absence of any long-range order for the unsaturated component of the spin within this proposed spin-nematic region. From our investigation we can claim the absence of any such signal within the $\vert \textbf{q} \vert$-range 0.062 to 0.30~\AA$^{-1}$, unless it lies outside the $(hk0)$ scattering plane, which we consider unlikely given the robustness of the [100] propagation direction of the spin-spiral observed throughout our measurements. Since a spin-nematic phase must preserve translational symmetry, these observations are consistent with this previously proposed interpretation of the high-field phase.

We observe the domain selection transition, which can be seen directly in figure~\ref{diff_pattern}(a,~b), in agreement with earlier measurements \cite{Fel12}. Here, domains with propagation vectors perpendicular to the magnetic field are removed, and the Bragg intensity in the selected domains increases rapidly, as seen in figure~\ref{intensity}(a,~b), as a result of the increase in volume fraction of the remaining domains. Furthermore, this transition is also seen in the length of the scattering vector, $\vert \textbf{q} \vert$. As the system lowers the number of magnetic domains, $\vert \textbf{q} \vert$ increases, which corresponds to a slight decrease in the length of the spin-spiral structure in direct space. In the helimagnetic compound MnSi, an application of pressure causes a shortening of the spin-spiral length \cite{Fak05}. Recalling that in ZnCr$_{2}$Se$_{4}$ the spin-spiral lies along the axis that undergoes the greatest distortion at $T_{\rm N}$ \cite{Kle66, Hid03}, these observations suggest that in the multi-domain state the structural distortion may induce local strains in the crystal lattice, thereby affecting the spin-spiral length, which are then released upon entering the single-domain state at $H_{\rm C1}$.

In general, both the magnetic structure and the phase transition lines are surprisingly insensitive to the direction of applied magnetic field. While there is a clear difference in the selection of a single domain for fields applied parallel to [100] and the selection of two domains for fields applied along [110], beyond this the magnetic field dependence for both orientations is nearly identical. Firstly, the direction of the propagation vector of the magnetic structure is rigid against varying magnetic field directions, only aligning along main crystallographic axes rather than following the field direction. Secondly, the phase diagram is identical regardless of the direction of applied field, with both the domain selection transition and the spin-spiral transition taking place at the same temperature and field for both $\textbf{H} \parallel [100]$ and $\textbf{H} \parallel [110]$. We illustrate this in figure~\ref{phase_diagram}. It is clear that the observed transitions fully coincide, within experimental uncertainties, irrespective of the direction of applied field. This is surprising, as one might expect the projection of the magnetic field on the direction of \textbf{q} to be important, which differs by a factor of $\sqrt{2}$ between the two experimentally chosen geometries. Even the domain selection transition, which is expected to be sensitive to different angles between applied field and the propagation vector in different domains, turns out to be fully isotropic in ZnCr$_{2}$Se$_{4}$. We note that this behaviour is in line with the observation that the magnetisation of ZnCr$_{2}$Se$_{4}$ also seems to be invariant with respect to the direction of the applied magnetic field \cite{Mur08}, indicating that the development of the conical spin structure proceeds at the same rate regardless of the projection of magnetic field onto the propagation direction of the spiral. We also see that the change of the propagation vector as a function of both field and temperature is similar for both field directions, with a reduction in $\vert \textbf{q} \vert$ with increasing field or temperature, as shown in figure \ref{q}. However, the magnetic structure appears to be more robust against this change in $\vert \textbf{q} \vert$ for fields applied parallel to the crystal axis than for the field applied along [110]. Therefore, while the precise magnetic structure does weakly depend on the magnetic field direction, the magnetic polarisation of the lattice and subsequent transitions do not.

\section*{Conclusions}

In summary, we have observed Bragg reflections from the helimagnetic structure in ZnCr$_{2}$Se$_{4}$, finding a systematic variation with temperature and field of the spin-spiral pitch length. Particularly, crossing the domain selection transition causes an abrupt jump in the propagation length of the magnetic structure of the order of a few percent. On approach with increasing field to $H_{\rm C2}$, we see a gradual order-parameter-like decrease in the diffracted neutron intensity in agreement with the observation that the screwlike magnetic structure transforms into a conical spiral as the spin lattice polarises under the application of field. Whilst previous magnetisation measurements have indicated that there still remains an unpolarised component of the spins within the intermediate field phase directly above $H_{\rm C2}$, we observe no magnetic Bragg reflections within this region, which is concurrent with the hypothesis that this state does not break the translational symmetry of the crystal, as anticipated for a spin-nematic phase.

\section*{Acknowledgements}
This work was supported by the German Research Foundation (DFG) through the Collaborative Research Centre SFB 1143 in Dresden (projects C01 and C03), individual research grant IN 209/3-1, and the Transregional Collaborative Research Centre TRR 80 (Augsburg, Munich, Stuttgart). This research was further supported by the Dresden High Magnetic Field Laboratory (HLD) at the 	 Helmholtz-Zentrum Dresden-Rossendorf, member of the European Magnetic Field Laboratory (EMFL).

\section*{References}
\bibliographystyle{iopart-num}
\bibliography{ZnCr2Se4}

\providecommand{\newblock}{}
\begin{thebibliography}{10}
\expandafter\ifx\csname url\endcsname\relax
  \def\url#1{{\tt #1}}\fi
\expandafter\ifx\csname urlprefix\endcsname\relax\def\urlprefix{URL }\fi
\providecommand{\eprint}[2][]{\url{#2}}
% Bibliography created with iopart-num v2.1
% /biblio/bibtex/contrib/iopart-num

\bibitem{Kle66}
Kleinberger R and de~Kouchkovsky R 1966 {\em C. R. Acad. Sci. Ser. B\/} {\bf
  262} 628

\bibitem{Plu66}
Plumier R~J 1966 {\em J. Appl. Phys.\/} {\bf 37} 964--965

\bibitem{Sir80a}
Siratori K, Akimitsu J, Kita E and Nishi M 1980 {\em J. Phys. Soc. Jpn.\/} {\bf
  48} 1111--1114

\bibitem{Mur08}
Murakawa H, Onose Y, Ohgushi K, Ishiwata S and Tokura Y 2008 {\em J. Phys. Soc.
  Jpn.\/} {\bf 77} 043709

\bibitem{Tok10}
Tokura Y and Seki S 2010 {\em Adv. Mater.\/} {\bf 22} 1554--1565

\bibitem{Kim12}
Kimura T 2012 {\em Annu. Rev. Condens. Matter Phys.\/} {\bf 3} 93--110

\bibitem{Fel12}
Felea V, Yasin S, G\"unther A, Deisenhofer J, Krug~von Nidda H~A, Zherlitsyn S,
  Tsurkan V, Lemmens P, Wosnitza J and Loidl A 2012 {\em Phys. Rev. B\/} {\bf
  86}(10) 104420

\bibitem{Hid03}
Hidaka M, Tokiwa N, Fujii M, Watanabe S and Akimitsu J 2003 {\em phys. stat.
  sol. (b)\/} {\bf 236} 9--18

\bibitem{Yok09}
Yokaichiya F, Krimmel A, Tsurkan V, Margiolaki I, Thompson P, Bordallo H~N,
  Buchsteiner A, St\"u\ss{}er N, Argyriou D~N and Loidl A 2009 {\em Phys. Rev.
  B\/} {\bf 79}(6) 064423

\bibitem{Aki78}
Akimitsu J, Siratori K, Shirane G, Iizumi M and Watanabe T 1978 {\em J. Phys.
  Soc. Jpn.\/} {\bf 44} 172--180

\bibitem{Cha90}
Chandra P, Coleman P and Larkin A~I 1990 {\em J. Phys.: Condens. Matter\/} {\bf
  2} 7933

\bibitem{Svi11}
Svistov L, Fujita T, Yamaguchi H, Kimura S, Omura K, Prokofiev A, Smirnov A,
  Honda Z and Hagiwara M 2011 {\em JETP Lett.\/} {\bf 93} 21--25

\bibitem{Fak05}
F\r{a}k B, Sadykov R~A, Flouquet J and Lapertot G 2005 {\em J. Phys.: Condens.
  Matter\/} {\bf 17} 1635

\end{thebibliography}

\end{document}